\documentclass[twocolumn,aps,showpacs]{revtex4}
\usepackage{graphicx}
\usepackage{dcolumn}
\usepackage{color}
\usepackage{amsmath}
\usepackage{bm}

\begin{document}

\title{Enhancement of photon intensity in forced coupled quantum wells
inside a semiconductor microcavity}
\author{ Hichem Eleuch$^{1,2}$, Awadhesh Prasad$^{1,3}$ and Ingrid Rotter$^1$}
\affiliation{$^1$Max Planck Institute for the Physics of Complex Systems,\\
N\"othnitzer Str.~38, D-01187 Dresden, Germany \\
$^2$Universite de Montreal, C.P. 6128, Succ. Centre---Ville, Montreal (QC),
H3C 3J7 Canada\\
$^3$Department of Physics and Astrophysics, University of Delhi, Delhi
110007, India}

\begin{abstract}
We study numerically the photon emission from a semiconductor microcavity
containing $N\ge 2$ quantum wells under the influence of a periodic external
forcing. The emission is determined by the interplay between external
forcing and internal interaction between the wells. While the external
forcing synchronizes the periodic motion, the internal interaction destroys
it. The nonlinear term of the Hamiltonian supports the synchronization. The
numerical results show a jump of the photon intensity to very large values
at a certain critical value of the external forcing when the number of
quantum wells is not too large. We discuss the dynamics of the system across
this transition.
\end{abstract}

\pacs{71.35.Gg, 05.45.-a, 05.45.Pq, 05.45.Ac }
\maketitle


\section{Introduction}

Over the past five decades nonlinear equations have been used extensively
for a detailed investigation of the optical properties of semiconductors
\cite{A01,A02,A03,A04,A05,Nad1} because of their potential application in
opto-electronic devices \cite{a6,a7,Nad2}. In semiconductor nanostructures
like quantum wells and quantum dots \cite{a4,a4b} the coupling between light
and matter is enhanced. By this, it may produce more pronounced nonlinear
and quantum effects such as a modification of the quantum statistical
properties of the emitted light as well as bistability and multistability.
These effects were theoretically predicted and experimentally observed by
several groups \cite{a11,a11b,a11c,a14,a15,a16,a17,a18,a19,a20}.

In natural systems as well as in experimental realizations of artificial
systems, the presence of an external forcing is unavoidable. An external
forcing could be either noise (from the surrounding or inherent within the
experiment) or caused by a deterministic perturbation. Sometimes, the
external forcing is useful for practical applications but sometimes it
entails some degradation of the desired system behavior. Examples of
important and useful results obtained by means of external perturbation are
stochastic resonances \cite{reso,reso1}, chaos control \cite%
{control,control1}, strange nonchaotic dynamics \cite{sna,sna1,sna2} etc.
One aspect to explore the nonlinear behavior is to scan the parameters space
and to observe how the dynamical complexity depends on the parameters \cite%
{tabor,kaneko,ruelle,schuster}.

The interaction between nonlinear systems gives rise to new phenomena such
as synchronization, hysteresis, phase locking, phase shifting, phase-flip,
riddling, amplitude death etc.~ \cite%
{synch,kanekoSych,riddle,hys,pfb,ap,pecora,pr}. Recently the coupled
nonlinear dynamical systems have been extensively studied from both the
theoretical and experimental points of view in a variety of contexts in the
physical, biological, and social sciences etc.~ \cite{synch,kanekoSych}.

In our previous work \cite{new} we explored the dynamics of the field
intensities in the high excitation regime inside a semiconductor microcavity
containing one quantum well. We observed periodic-doubling, quasiperiodic
and direct route to chaos as the forcing strength is changed. These results
show various types of dynamics depending on the forcing strength.
Furthermore we observed also coexisting periodic and chaotic attractors with
riddled basin.

In the present paper we consider a network of quantum wells as shown
schematically in Fig. 1\,: $N$ quantum wells are inside a semiconductor
microcavity (schematically represented by two Bragg mirrors $M$) in the
presence of an external forcing. The aim is to analyze the dynamical
behavior of the intra-cavity photonic intensity and the intensity of the
fluorescent light in the presence of periodic signals. We observed a jump in
the intensities of photon and exciton emission at appropriate values of the
forcing strength. The maximum of the photon intensity is obtained for an
optimal number of quantum wells which should not be too large. Across this
transition we see either periodic or chaotic motion depending upon the
exciton resonance frequency of the individual quantum wells.

The paper is organized as follows. In Sec. II, we review the derivation of
exciton-phonon interaction in the presence on $N$ quantum wells. This is
followed by Sec. III where the results for several quantum wells are
considered. The results are discussed and some conclusions are drawn in Sec.
IV.

\section{Model}

\begin{figure}[tbp]
\includegraphics[scale=.4]{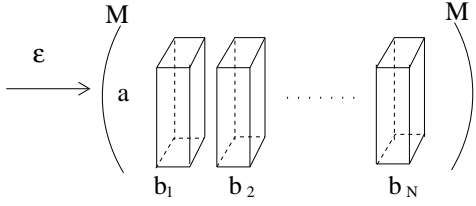}
\caption{ The schematic model for microcavity with $N$-quantum wells.
Symbols are discussed in text.}
\label{sch}
\end{figure}

We consider $N$ quantum wells inside a semiconductor microcavity, see Fig.
1. Each quantum well is localized in a position, which corresponds to the
maximum of the electromagnetic field inside the microcavity. This system is
similar to the one described in detail in \cite{a22}. Neglecting the effects
of the spins, the interaction with the phonons and the excitonic saturation,
we can write the Hamiltonian describing the system with N quantum wells
inside the cavity and pumped with a laser with amplitude $\varepsilon $ and
frequency $\omega _{L}$ as \cite{a23,a24,a241,A23,A24,A25,A26}:
\begin{eqnarray}
H &=&\hbar \omega _{ph}a^{+}a+\underset{j=1}{\overset{N}{\sum }}\left[ \hbar
\omega _{j}b_{j}^{+}b_{j}\right] +i\hbar \varepsilon \left( a^{+}e^{-i\omega
_{L}t}-ae^{i\omega _{L}t}\right)   \notag \\
&&+\underset{j=1}{\overset{N}{\sum }}\left[ i\hbar g_{j}^{^{\prime
}}\left(
a^{+}b_{j}-b^{+}a_{j}\right) +\hbar \alpha _{j}^{^{\prime
}}b_{j}^{+}b_{j}^{+}b_{j}b_{j}\right]   \label{h1}
\end{eqnarray}%
The first two terms of the Hamiltonian correspond to the proper energies of
photon and quantum well-excitons, where $a$, $b_{j}$ are respectively the
annihilation operators of a photonic and excitonic modes verifying: $\left[
a,a^{+}\right] =1$ and $\left[ b_{j},b_{j}^{+}\right] =1.$ $\omega _{ph}$ is
the photonic mode frequency and $\omega _{j}$ is the excitonic mode
frequency for the $j^{th}$ quantum well. The third term corresponds to the
pump energy. The first part of the fourth term
represents the exciton-photon coupling with { a coupling constant $g_{j}^{^{\prime
}}$}. The
last part of the fourth term describes the excitonic nonlinearity for each
quantum well with the coefficients $\alpha _{j}^{^{\prime }}.$

By considering that the fluctuations are weak compared to the
average values, the evolution of the mean field operators in the interaction picture can be written as:
\begin{subequations}
\label{model}
\begin{align}
\frac{d\left\langle a\right\rangle }{d\tau }& =\epsilon
(t)+\sum_{j=1}^{N}g_{j}\left\langle b_{j}\right\rangle -\kappa \left\langle
a\right\rangle -i\Delta _{a}\left\langle a\right\rangle \\
\frac{d\left\langle b_{j}\right\rangle }{d\tau }& =-g_{j}\left\langle
a\right\rangle -\frac{\gamma _{j}}{2}\left\langle b_{j}\right\rangle
-i\Delta _{j}\left\langle b_{j}\right\rangle -2i\alpha _{j}\left\langle
b_{j}^{+}\right\rangle \left\langle b_{j}\right\rangle \left\langle
b_{j}\right\rangle
\end{align}%
where $j=1,...,N$,  ~$\tau ~$ is a dimensionless time normalized
  to $\tau_{c}=\frac{3}{2g_{1}}:$
\end{subequations}
\begin{equation}
\tau =\frac{t}{\tau _{c}} \; ,
\end{equation}%
$\gamma _{j},~\kappa $ are the dimensionless decay rates of the excitons and
the cavity photon:
\begin{equation}
\gamma _{j}=\gamma _{ex_{j}~}\tau _{c};\kappa =\gamma _{ph~}\tau _{c}
\; ,
\end{equation}%
{ the nonlinear coupling constant $\alpha_{j}$ and the coupling $g_j$ are normalized to $\frac{1}{\tau _{c}}:$
\begin{eqnarray}
\alpha _{j}=\alpha _{j}^{^{\prime }}\tau _{c}  \notag \\ 
g_{j}=g_{j}^{^{\prime }}\tau _{c}, 
\end{eqnarray}%
}
and $\Delta _{a},$ $\Delta _{j}$ are the dimensionless detuning
\begin{eqnarray}
\Delta _{a} &=&\left( \omega _{ph}-\omega _{L}\right) \tau _{c}  \notag \\
\Delta _{j} &=&\left( \omega _{j}-\omega _{L}\right) \tau _{c}.
\end{eqnarray}%
 According to these equations, the coupling among excitonic modes
  arises from their common interaction with the photonic mode.
It represents  a global mean field coupling of excitonic modes and is
given by the sum term (over $j$) in (\ref{model}). Moreover, the evolution of
the excitonic modes contains a nonlinear term (last term in  (\ref{model}).
{ It is worth to mention that the nonlinear parameters $\alpha_j$ can be scaled to 1 by redefining $<b_j^{new}>=<b_j>/\sqrt\alpha_j$  and $g_j^{new}=g_j*\sqrt\alpha_j$.}
This system has $2(N+1)$-dimensions in the presence of forcing . For the
numerical simulation we use RK4 integrator \cite{nr}. We consider the step
size $\Delta t=2\pi /5000$ for integration. The dynamical studies are
studied after removing initial $10^{7}$ data points as transients. We
explore here the photon and exciton intensities $I_{a}=$ $\left\langle
a^{+}\right\rangle \left\langle a\right\rangle $ and $I_{j}=\left\langle
b_{j}^{+}\right\rangle \left\langle b_{j}\right\rangle $ inside the cavity.
As the fluorescent light is proportional to the mean number of excitons ($%
I_{j}$), we are also exploring the fluorescent light dynamics.

\section{Numerical results}

\label{res}

The system properties are determined by different physical parameters. In
our calculations, we fix some of them in order to have the possibility to
determine the influence of the remaining ones. In all our calculations, the
normalized parameters are fixed to $g_{j}=g=1.5$ and $\kappa =0.12$, $\gamma
=\gamma _{j}=0.015$ which correspond to the experimental values \cite{a18}
in units of the inverse of the round-trip in the microcavity. Furthermore,
we concentrate our analysis to the case of a strong pump field where the
non-linear phenomena are expected to influence the dynamics. We choose the
normalized amplitude of the laser pump as $\epsilon =200$ and the normalized
excitonic interaction coefficient as $\alpha_{j}=\alpha =0.00001$. At this
set of parameters the equations describing the dynamics of the system have
stable fixed point solutions.

The parameter $\epsilon $ characterizes the influence of external forcing on
the dynamics of the system. We use it, in the present study, in order to
receive information on the effect of the external forcing. We restrict our
analysis to a deterministic periodic forcing in $\epsilon $, i.e. $\epsilon
\to \epsilon (1+f\, cos(\Omega t))$ where $\Omega $ is the perturbation
frequency while $f$ is the strength of the forcing. The last value is
considered as bifurcation parameter. As to the first value, we consider only
the case $\Omega =1$ in the present work. { This value of $\Omega$ corresponds to 1.5 Thz physical frequency.
The Thz-sources are recently realized \cite {50,51}.}

As we change the magnitude of $f$, various dynamical motions are possible.
We have discussed this result recently \cite{new} for the case of one
quantum well and different values of the detuning $\Delta _{a}$ and $\Delta
_{1}$. In the present paper we consider the effect of the forcing strength $%
f $, of the detuning $\Delta _{j}$ and of the number $N$ of quantum wells at
fixed $\Delta _{a}=0$ (the cavity is resonant with the pump laser).

\subsection{Two quantum wells}

\label{two}

In order to show the results in details, we first consider two quantum
wells. We fix $\Delta _{1}=-g$ and vary $\Delta _{2}\in[-g,g]$.

\begin{figure}[tbp]
\includegraphics[scale=.7]{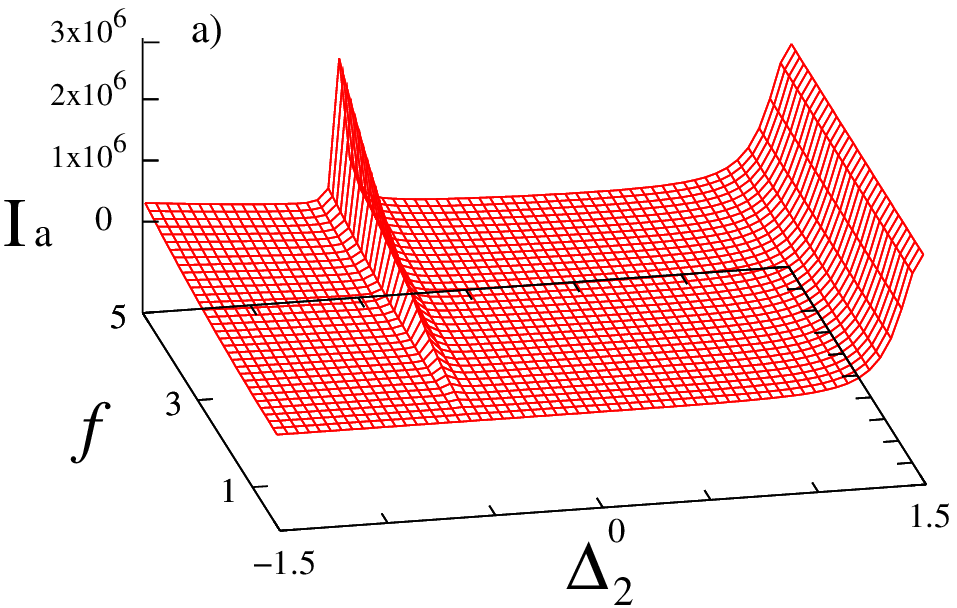}
\includegraphics[scale=.7]{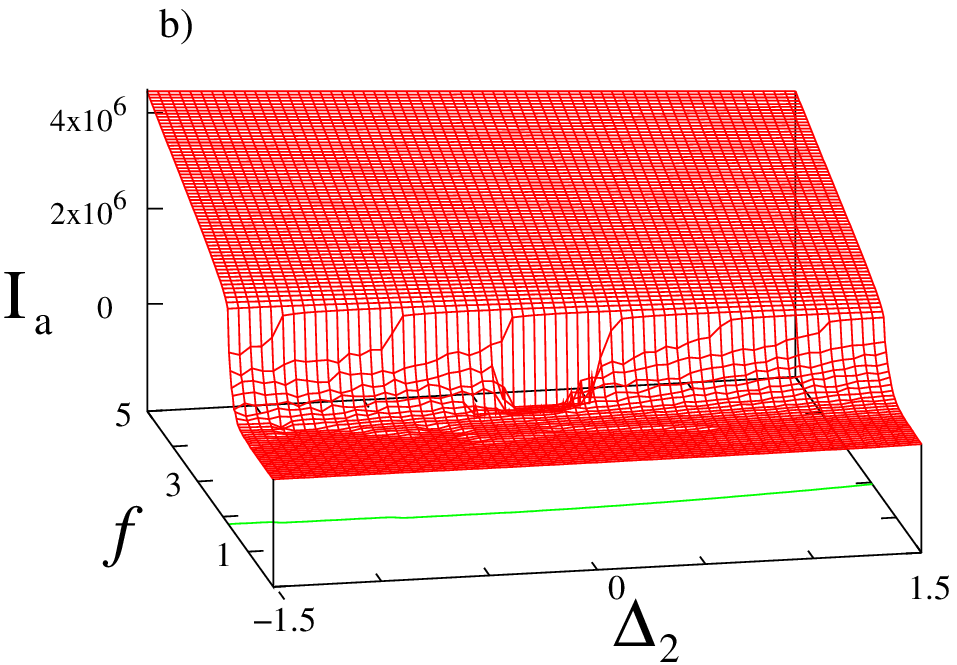}
\caption{The photon intensity $I_a $ in the parameter space $\Delta_{2} - f$
at the fixed value $\Delta_1=-g$ for two quantum wells without (a) and with
(b) nonlinear term in the Hamiltonian Eq. (\protect\ref{h1}). The contour
line in (b) is drawn at $I_a=2\times 10^6$.}
\label{d1d2}
\end{figure}

Shown in Figs. \ref{d1d2}(a,b) are the photon intensities $I_a$ in the
parameter space $\Delta _{2} - f$, respectively, without ($\alpha_j=0$) and
with ($\alpha_j=0.00001$) a nonlinear term in the Hamiltonian (\ref{h1}).
Fig. \ref{d1d2}(a) shows that $I_a$ varies smoothly as a function of the
forcing $f$ except near to the value of $\Delta_2=-0.5$ when the nonlinear
term is not considered. The results obtained with inclusion of the nonlinear
term show another behavior, Fig. \ref{d1d2}(b). Here, the intensity
increases smoothly as a function of $f$ for all but a certain critical value
of $f$. At this critical value, $I_a$ jumps to much higher values.

\begin{figure}[tbp]
\includegraphics[scale=0.4]{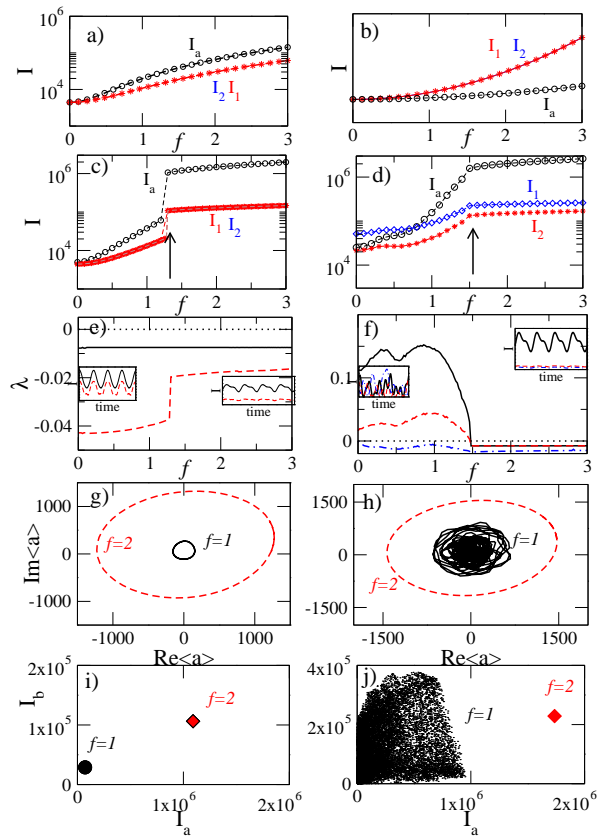}
\caption{ (Color online) Comparison of some results for two identical
quantum wells with those for two mismatched wells. The left and right panels
correspond to $\Delta_1=\Delta_2=g$ and $\Delta_1=-g,\Delta_2=g$,
respectively. (a-d): Intensities $I_a $ ($\circ$), $I_{1}$ ($\diamond$), and
$I_{2}$ ($\star$) as a function of the forcing strength $f$ without (a,b)
and with (c,d) nonlinear term in (\protect\ref{h1}). The synchronized
intensities $I_{1}$ ($\diamond$) and $I_{2}$ ($\star$) in (a) and (c) are
overlapping. (e,f): A few largest Lyapunov exponents as a function of the
forcing strength $f$. The dotted lines represent the zero Lyapunov exponent.
Insets in (e,f): intensities $I_a $ (black-solid line), $I_1$ (red-dashed
line) and $I_2$ (blue-long-dashed line) as a function of time corresponding
to (g,h). (g,h): Trajectories in the phase space $Im \langle a\rangle
-Re\langle a\rangle$ at $f=1$ (inner solid line) and $f=2$ (outer red-dashed
line) below and above, respectively, the transition. (i,j): The Poincare
section of (g,h). The other parameters are the same as those in Fig. 2. }
\label{d1d2-g}
\end{figure}

The details of the behavior of $I_{a}$ are shown in Fig. \ref{d1d2-g} where
the left panel is drawn for fixed $\Delta _{1}=\Delta _{2}=g$ while the
right panel corresponds to $\Delta _{1}=-g$ and $\Delta _{2}=g$. These two
cases correspond to identical and mismatched quantum wells, respectively.
The mismatch in the latter case is equal to the Rabi frequency of the single
quantum well which is $|\Delta _{1}-\Delta _{2}|=2g$.

The results shown in Figs. \ref{d1d2-g}(a,b) are obtained with $\alpha_j=0$,
i.e. with a vanishing nonlinear term in the Hamiltonian $H$, Eq. (\ref{h1}).
They show an overall smooth increase of the intensity with increasing
forcing strength $f$. The results in Figs. \ref{d1d2-g}(c,d) are obtained
with a nonvanishing nonlinear term in (\ref{h1}), i.e.~ with $\alpha
_{j}=\alpha =0.00001$. In this case, $I_a$ does not increase everywhere
smoothly with increasing forcing $f$, see also Fig. \ref{d1d2}(b). The
nonlinearity causes a substantial jump in the intensity around $f\sim 2$.
This jump occurs independently of the detuning $\Delta _{2}$. In Figs. \ref%
{d1d2-g}(c,d) the jump is marked by an arrow.

In order to see the dynamical behavior of the system across this transition
we plot a few of the largest Lyapunov exponents (LEs) in Figs. \ref{d1d2-g}%
(e,f) for the nonlinear cases considered in Figs. \ref{d1d2-g}(c) and (d),
respectively. The LEs are calculated according to \cite{lapun}. In the case
of identical quantum wells (left panel) the dynamics is always periodic and
all but one Lyapunov exponent (which distinguish the type of dynamics, see
Ref. \cite{tabor}) are negative. One of the LEs is zero (dotted line). The
third LE (dashed line) is negative but jumps at the same value $f=1.3$ at
which the intensity jumps (shown by an arrow in Fig. \ref{d1d2-g}(c)). The
trajectories in the phase space across this transition are shown for two
different values of the forcing strength $f$ in Fig. \ref{d1d2-g}(g). The
inner black solid line corresponds to $f=1$ while the outer red-dashed line
is for $f=2$. These results are confirmed by the corresponding Poincar\'{e}
sections \cite{tabor} which are taken at $Re\langle a\rangle =0$, see Fig. %
\ref{d1d2-g}(i). There are two single points corresponding to $f=1$ and $2$.

It should be noted here that both the photon intensity and the exciton
intensity jump at the same values of $f$. The photon intensity is, however,
much larger than the exciton intensity, see Fig. \ref{d1d2-g}(c). The
variation as a function of time is shown in the insets of Fig. \ref{d1d2-g}%
(e) which are calculated with the parameters of Fig. \ref{d1d2-g}(g).

In the mismatched case (right panel) we see a behavior of the intensities $I$
as a function of the forcing strength $f$ which is similar to that discussed
above for the case with identical wells. Due to the mismatching, the
critical values differ from those of the left panel. However the exciton
intensities are much smaller than the photon intensities also in this case.
The dynamics across the transition is shown in Fig. \ref{d1d2-g}(f). The
spectrum of the LEs shows the following characteristic features\,: below the
transition, the motion is chaotic while it is periodic beyond the
transition. Both, the chaotic and the periodic motion are shown in Fig. \ref%
{d1d2-g}(h) with black-solid and red-dashed lines, respectively, at the
forcing strengths $f=1$ and $2$. These results are confirmed by the
corresponding Poincar\'{e} sections taken at $Re\langle a\rangle =0$ and
shown in Fig. \ref{d1d2-g}(j). The motion is chaotic (scattered points) and
periodic (single point), respectively. The variation of the intensities as a
function of time is shown in the insets of Fig. \ref{d1d2-g}(f) at the
parameter values $f=1$ and $2$.

Comparing the results obtained for the case with two identical quantum wells
(left panel of Fig. \ref{d1d2-g}) to those obtained with two mismatched
wells (right panel of Fig. \ref{d1d2-g}), we state the following. In the
first case, the intensities jump at a certain critical value of the forcing
strength $f$ while the transition starts much below this critical value of $%
f $ in the second case. Note that the y-axis is taken on logarithmic scale.
Further, there is a synchronized periodic motion (where the $I_j$ have
identical values \cite{synch}) across the jump in the first case with
identical wells while the motion changes from an unsynchronized chaotic to a
synchronized periodic one in the second case with mismatched wells. In any
case, the intensities $I$ at large $f$ are much larger when the excitonic
nonlinearity in the Hamiltonian Eq. (\ref{h1}) is taken into account, than
without this term. The nonlinearity causes, obviously, the jump-like
transition to the higher intensities at the critical value of the forcing
strength $f$.

\subsection{Chain of N quantum wells}

\label{more}

\begin{figure}[tbp]
\includegraphics[scale=.55]{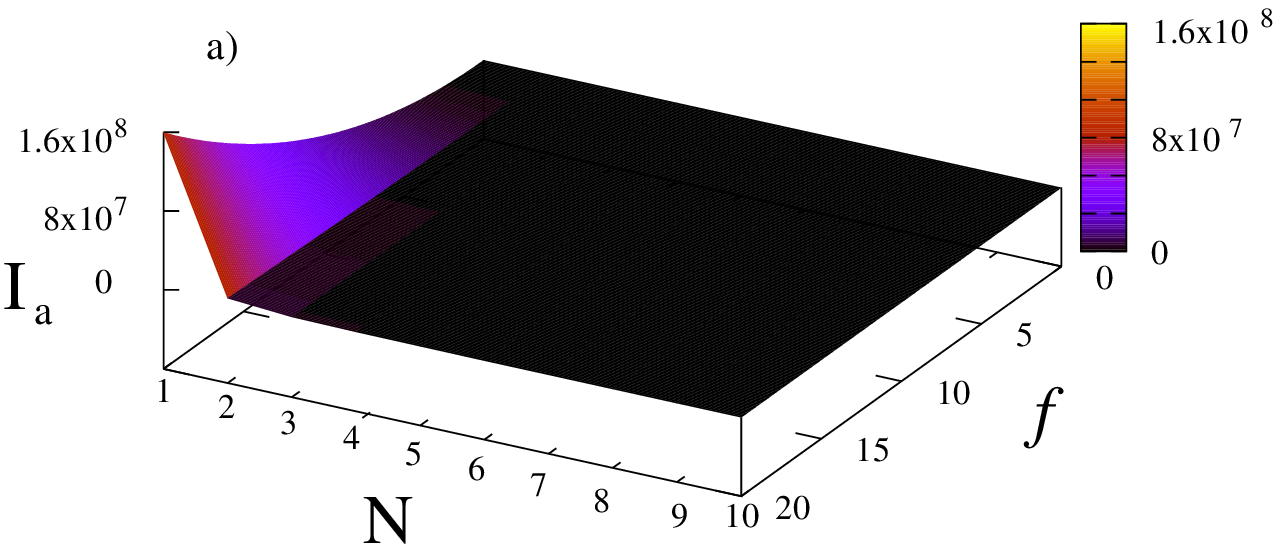} %
\includegraphics[scale=.55]{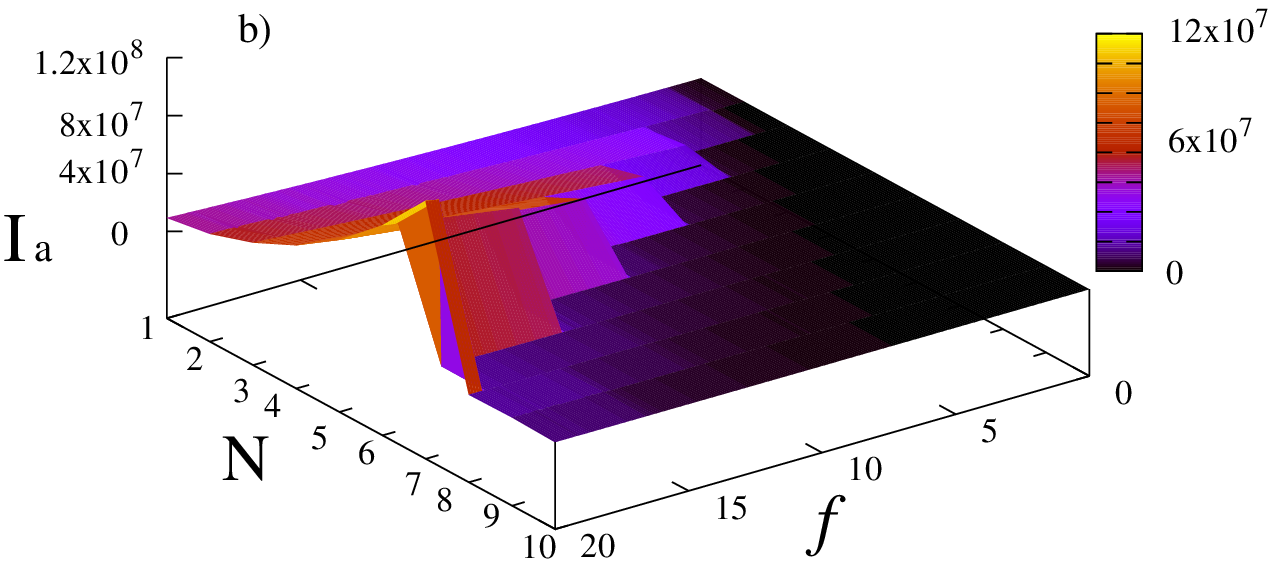}
\caption{(Color online) The photon intensity $I_a $ in the parameter space $%
f-N$ for $N$ identical quantum wells at fixed $\Delta_j=g$, $\forall j$
without (a) and with (b) nonlinear term in (\protect\ref{h1}). }
\label{fn}
\end{figure}

In order to see the effect of the nonlinearity on the photon and exciton
intensities in the case of a large number $N$ of quantum wells we consider
in the following a chain of identical quantum wells with fixed $\Delta_j=g,
~\forall j$. In Fig. \ref{fn} the calculated photon intensity $I_a$ is drawn
in the parameter space~: number $N$ of quantum wells and forcing
strength $f$.

According to Fig.~\ref{fn}(a) the intensity $I_a$ decreases smoothly with
increasing $N$ ($>2$) when the nonlinear term in (\ref{h1}) in not taken into
account (corresponding to $\alpha_j =0$). However there is a jump in
intensity from $N=1$ to $N=2$. Fig. \ref{fn}(b) shows the results
obtained with inclusion of the nonlinear term in (\ref{h1}), i.e. with $%
\alpha_j \ne 0$. In this case, the intensity increases first with increasing
$N$ and then jumps to lower values. This behavior repeats several times.
Both figures indicate further that, at very low forcing strength $f\sim 0$, ~%
$I_a$ does not change for any value of $N$. As a result, a jump in the
intensity $I_a$ appears only when the forcing strength $f$ as well as the
nonlinearity $\alpha$ do not vanish.

\begin{figure}[tbp]
\includegraphics[scale=0.4]{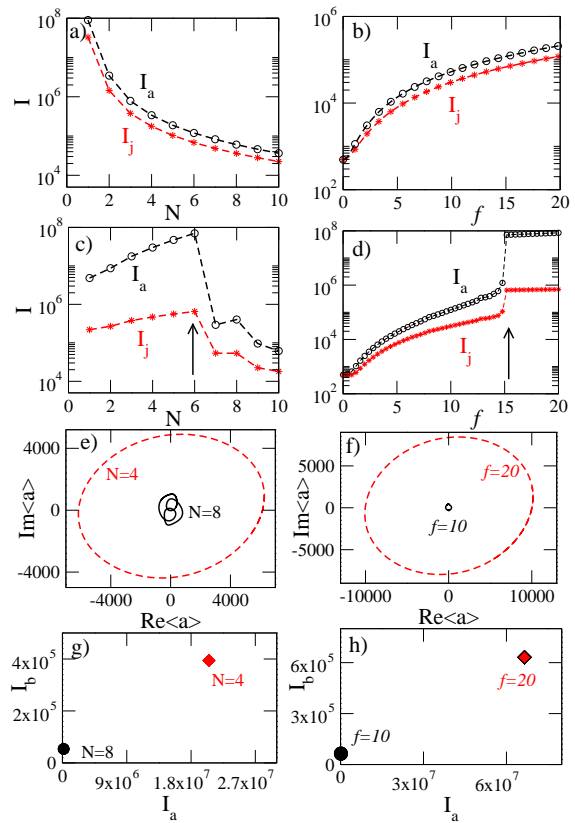}
\caption{ (Color online) Some results for $N$ identical quantum wells as a
function of the number $N$ (at fixed $f$, left panel) and of the forcing
strength $f$ (at fixed $N$, right panel). (a-d): Intensities $I_a $ ($\circ$%
) and $I_{j}$ ($\star$) as a function of $N$ at fixed forcing strength $f=15$
(a,c) and as a function of $f$ at fixed number $N=6$ of quantum wells (b,d).
The results are obtained, respectively, without (a,b) and with (c,d)
nonlinear term in (\protect\ref{h1}). (e,f): Trajectories in phase space $Im
\langle a\rangle - Re\langle a\rangle $ at (e) $N=4$ (outer red-dashed line)
and $N=8$ (inner black solid line) for fixed $f=15$; and at (f) $f=10$
(inner solid line) and $f=20$ (outer red-dashed line) for fixed $N=6$. (g,h)
The Poincare section of (e,f). The other parameters are the same as those in
Fig. 4.}
\label{fn-n-f}
\end{figure}

When the nonlinearity in (\ref{h1}) is taken into account in the
calculations, a substantial drop in the intensity $I_a$ appears at some
values of the forcing strength $f$ when we increase the number $N$ of
quantum wells. This behavior of $I_a$ is determined obviously by the
nonlinear term in (\ref{h1}). The details of the variation of the photon
intensity $I_a$ as well as of the exciton intensity $I_j$ (all $I_j$ are
synchronized) are shown in Fig. \ref{fn-n-f}.

Figs. \ref{fn-n-f}(a, b) show numerical results obtained without the
nonlinearity in (\ref{h1}) as a function of $N$ and $f$, respectively. Here,
the intensities vary smoothly~: the photon intensity $I_a$ as well as the
exciton intensity $I_j$ decrease with increasing $N$ but increase with
increasing $f$. The situation is, however, completely different when the
nonlinearity is taken into account. In this case, substantial jumps appear
in the intensities, see Figs. \ref{fn-n-f}(c,d) and the further results
shown in Figs. \ref{fn-n-f}(e-h).

Let us first consider the variation of $I_a$ as a function of $N$ for fixed
forcing strength $f=15$. According to Fig. \ref{fn-n-f}(c), $I_a$ increases
first with the number $N$ of quantum wells. However, it reaches a maximum
value at $N=6$ where it decreases suddenly and drops to very low values
(even lower than in the single quantum well). The details of the dynamics
across this transition are shown in Fig. \ref{fn-n-f}(e). Here, trajectories
at $N=4$ (outer red-dashed line) and $8$ (inner black-solid line) are drawn
in the phase space $Re\langle a \rangle-Im\langle a\rangle$. These
trajectories show periodic motions what is confirmed by the Poincare section
shown in Fig. \ref{fn-n-f}(g) with points corresponding to $N=4$ and $8$.

Fig. \ref{fn-n-f}(d) shows the intensity $I_a$ as a function of the forcing
strength $f$ for a fixed number $N=6$ of quantum wells. Similar to the case
with two identical wells (Fig. \ref{d1d2-g}(c)) the intensity $I_a$
increases smoothly up to a certain value of $f$ (marked by an arrow in the
figure) where it jumps to a much larger value. The corresponding dynamics in
phase space $Re\langle a\rangle -Im\langle a\rangle $ is shown in Fig. \ref%
{fn-n-f}(f) for $f=10$ (inner black solid line) and $f=20$ (outer red-dashed
line). Both motions are periodic. This result is confirmed by the Poincare
section, see Fig. \ref{fn-n-f}(h) where the two corresponding points $f=10$
and $20$ are shown.

The exciton intensities $I_j$ show a similar behavior as the photon
intensity $I_a$ in all cases. They are however smaller than $I_a$.

The results shown in Fig. \ref{fn-n-f} indicate that the dynamics of the
system consisting of $N$ forced coupled semiconductor microwave cavities is
determined by two opposite tendencies. On the one hand, the interaction
between the quantum wells prevents a synchronized periodic motion. On the
other hand, however, the nonlinear terms in the Hamiltonian (\ref{h1})
support the synchronized motion. As a result of these two tendencies, it is
possible to increase the photon intensity $I_a$ substantially when the
number of quantum wells is not too large.

\section{Discussion and conclusions}

\label{disc}

In Sect. \ref{res}, we showed the calculated intensities of photon and
exciton emission from a semiconductor microcavity containing of $N$ quantum
wells under the influence of an external forcing $f$. The intensity of the
photon and exciton emission, $I_a$ and $I_j$ respectively, is determined by
the interplay of external forcing and internal interaction between the
single quantum wells of the microcavity. External forcing synchronizes the
periodic motions and causes, by this means, an enhancement of the
intensities. The internal interaction, however, disturbs the synchronized
motion and leads to a reduction of the intensities of photon as well of
exciton emission. The interplay between external and internal interaction is
described well by the Hamiltonian (\ref{h1}).

In the case of $N=2$, the destructive role of the internal interaction is
small. The intensities $I_a$ and $I_j$ increase with $f$ starting at a
certain small value of $f$. This holds true for the case of two identical
wells as well as for two mismatched wells. In both cases, the intensities
jump to much higher values at a certain critical value of the forcing
strength $f$. In the first case, the dynamics of the system is a
synchronized periodic motion below as well as beyond the jump. In the second
case, however, the motion is unsynchronized chaotic below the critical value
of $f$ where the jump occurs, and changes to a synchronized periodic motion
for $f$ values beyond the jump.

In the case with $N$ wells, the destructive role of the internal interaction
between the wells can directly be seen. Without the nonlinear term in the
Hamiltonian (\ref{h1}), the intensities decrease with increasing $N$ (and
fixed $f$) but increase with $f$ (and fixed $N$), see Figs. \ref{fn-n-f}%
(a,b). With nonlinear term in (\ref{h1}), however, the intensities increase
first smoothly up to a certain critical value of $N$ (and fixed $f$), and a
critical value of $f$ (and fixed $N$), respectively, see Figs. \ref{fn-n-f}%
(c,d). Beyond these values, the intensities decrease abrupt in the first
case (Fig. \ref{fn-n-f}(c)) while they jump to higher values in the second
case (Fig. \ref{fn-n-f}(d)). In both cases, the intensity $I_a$ jumps by
several orders of magnitude at the critical points.

These results illustrate very nicely the interplay of internal interaction
between the wells and external forcing. The jump in the intensities is a
coherent effect that does not occur in a single quantum well, see the
results obtained earlier \cite{new}. The jump in the intensities with a
fixed (not too large number $N$ of wells) at a critical value of $f$ is
similar to that observed in the case with two wells. In any case, the
dynamics across the jump depends on the internal parameters of the quantum
wells, and the nonlinear term in the Hamiltonian (\ref{h1}) plays an
important role.

The enhancement of the intensity of photons emitted from a microcavity is of
great interest for applications. According to the results of the present
paper it is possible to manipulate microcavities in such a way that the
intensity of emitted light is very large.

\vspace{.5cm}

H.E. and A.P. thank E. Siminos for valuable comments and acknowledge the
financial support and the hospitality of the MPIPKS.


\vskip1cm

\begin{thebibliography}{99}
\bibitem{A01} S. Schmitt-Rink, D. A. B. Miller and D. S. Chemla, Phys. Rev.
B \textbf{35}, 8113 (1997).

\bibitem{A02} G. Khitrova, H. M. Gibbs, F. Jahnke, M. Khira and S. W. Koch,
Rev. Mod. Phys. \textbf{71}, 1591 (1999).

\bibitem{A03} V. M. Axt and S. Mukamel, Rev. Mod. Phys. \textbf{70}, 145
(1989).

\bibitem{A04} C. T. Sah, L. Forbes, L. L. Rosier and Jr. A. F. Tash,
Solid-State Electronics \textbf{13}, 759 (1970).

\bibitem{A05} E. Garmire and A. Kost, \textsl{Nonlinear Optics in
Semiconductors I} (Academic Press, London, 1999).

\bibitem{Nad1} N. Boutabba, L. Hassine, A. Rihani and H. Bouchriha, Synth.
Met. \textbf{4}, 227 (2003).

\bibitem{a6} T. C. H. Liew, I. A. Shelkhy and G. Mapluech, Physica E \textbf{%
43}, 1543 (2011).

\bibitem{a7} A. Amo, T. C. H. Liew, C. Adrados, R. Houdr\'e, E. Giacobino,
A. V. Kavokin and A. Bramati, Nat. Photon. \textbf{4}, 361 (2010).

\bibitem{Nad2} N. Boutabba, L. Hassine, N. Loussaief, F. Kouki and H.
Bouchriha, Organic Electronics \textbf{4}, 1 (2003).

\bibitem{a4} G. Khitrova, H. M. Gibbs, M. Khira, S. W. Koch and A. Scherer,
Nature Phys. \textbf{2}, 81 (2006).

\bibitem{a4b} T. Yoshie, A. Scherer, J. Hendrickson, G. Khitrova, H. M.
Gibbs, G. Rupper, C. Ell, O. B. Shchekin and D. G. Deppe, Nature \textbf{432}%
, 200 (2004).

\bibitem{a11} B. Deveaud, \textsl{The Physics of Semiconductor Microcavities}
(Wiley-VCH, New York, 2007).

\bibitem{a11b} A. Quattropani and P. Schwendimann, Phys. Stat. Sol. B
\textbf{242}, 2302 (2005).

\bibitem{a11c} E. Giacobino, J. P. Karr, A. Baas, G. Messin, M. Romanelli
and A. Bramati, Solid Stat. Commun. \textbf{134}, 97 (2005).

\bibitem{a14} E. Giacobino, J. P. Karr, G. Messin, H. Eleuch, A. Baas, C. R.
Physique \textbf{3}, 41 (2002).

\bibitem{a15} G. Messin, J. P. Karr, H. Eleuch, J. M. Courty and E.
Giacobino, J. Phys.: Condens. Matter \textbf{11}, 6069 (1999).

\bibitem{a16} H. Eleuch, J. M. Courty, G. Messin, C. Fabre and E. Giacobino,
J. Opt. B.: Quantum Semiclass.Optics \textbf{1}, 1 (1999).

\bibitem{a17} J. P. Karr, A. Baas, R. Houdr\'e, and E. Giacobino, Phys. Rev.
A. \textbf{69}, 031802 (2004).

\bibitem{a18} A. Baas, J. P. Karr, H. Eleuch, and E. Giacobino, Phys. Rev. A
\textbf{69}, 023809 (2004).

\bibitem{a19} T. K. Paraso, M. Wouters, Y. L\'eger, F. Morier-Genoud and B.
Deveaud-Pl\'edran, Nature Materials \textbf{9}, 655 (2011).

\bibitem{a20} E. A. Cotta and F. M. Matinaga, Phys. Rev. B. \textbf{76},
073308 (2007).

\bibitem{reso} R. Benzi, A. Sutera and A. Vulpiani, J. Phys. A \textbf{14},
L453 (1981).

\bibitem{reso1} C. Nicolis and G. Nicolis, Scholarpedia, \textbf{2(11)},
1474 (2007).

\bibitem{control} A. C\'ordoba, M. C. Lemos, and F. Jim\'enez-Morales, J.
Chem. Phys. \textbf{124}, 014707 (2006).

\bibitem{control1} I. Z. Kiss and J. L. Hudson, Phys. Rev. E \textbf{64},
046215 (2001).

\bibitem{sna} C. Grebogi, E. Ott, S. Pelikan and J. A. Yorke, Physica D
\textbf{13}, 261 (1984).

\bibitem{sna1} A. Prasad, S. S. Negi, and R. Ramaswamy, Int. J. Bif. and
Chaos \textbf{11}, 291 (2001).

\bibitem{sna2} A. Prasad, A. Nandi and R. Ramaswamy, Int. J. Bif. and Chaos
\textbf{17}, 3397 (2007).

\bibitem{tabor} M. Tabor, \textsl{Chaos and Integrability in Nonlinear
Dynamics: An Introduction} (Wiley, New York, 1989).

\bibitem{kaneko} K. Kaneko, \textsl{Collapse of tori and genesis of chaos in
dissipative systems} (Wold Scientific Publication, Singapore, 1986).

\bibitem{ruelle} D. Ruelle and F. Takens, Commun. Math. Phys. \textbf{20},
167 (1971).

\bibitem{schuster} H. G. Schuster and W. Just, \textsl{Deterministic Chaos:
An Introduction} (Wiley-VCH Weinheim, 2005).

\bibitem{synch} A. S. Pikovsky, M. G. Rosenblum, and J. Kurths, \textit{%
Synchronization: A Universal Concept in Nonlinear Sciences} (Cambridge
University Press, Cambridge, U.K., 2001).

\bibitem{kanekoSych} K. Kaneko, \textsl{Theory and Applications of Coupled
Map Lattices} (John Wiley and Sons, New York, 1993).

\bibitem{riddle} J. C. Alexander, J. A. Yorke, Z. You, and I. Kan, Int. J.
Bif. Chaos \textbf{2}, 795 (1992).

\bibitem{hys} A. Prasad, L. D. Iasemidis, S. Sabesan and K. Tsakalis,
Pramana, J. Phys. \textbf{64}, 513 (2005).

\bibitem{pfb} A. Prasad, J. Kurths, S. K. Dana, and R. Ramaswamy, Phys. Rev.
E. \textbf{74}, 035204 (2006).

\bibitem{ap} A. Prasad, Phys. Rev. E \textbf{72}, 056204 (2005).

\bibitem{pecora} L. M. Pecora and T. L. Carroll, Phys. Rev. Lett. \textbf{64}%
, 821 (1990).

\bibitem{pr} G. Saxena, A. Prasad and R. Ramaswamy, Physics Reports,
(DOI:10.1016/j.physrep.2012.09.003)-- in press.

\bibitem{new} H. Eleuch and A. Prasad, Phys. Lett. \textbf{376}, 1970 (2012).

\bibitem{a22} R. Houdre, C. Weisbuch, R. P. Stanley, U. Oesterle and M.
Ilegems, Phys. Rev. Lett. \textbf{85}, 2793 (2000).

\bibitem{a23} H. Haug, Z. Phys. B. \textbf{24}, 351 (1976).

\bibitem{a24} E. Hanamura, J. Phys. Soc. Jpn. \textbf{37}, 1545 (1974).

\bibitem{a241} E. Hanamura, J.Phys.Soc.Jpn. \textbf{37}, 1553 (1974).

\bibitem{A23} E. A. Sete, H. Eleuch and S. Das, Phys. Rev. A. \textbf{84},
053817 (2011).

\bibitem{A24} H. Eleuch and N. Rachid, Eur. Phys. J. D \textbf{57}, 259
(2010).

\bibitem{A25} H. Eleuch, Applied Mathematics \& Information Sciences \textbf{%
3}, 185 (2009).

\bibitem{A26} H. Eleuch J. Phys. B: At. Mol. Opt. Phys. \textbf{41}, 055502
(2008).

\bibitem{50} D. Shrekenhamer et al., Optics Express \textbf{19}, 9968 (2011).
\bibitem{51} S. Busch et al., Opt. Lett. \textbf{37}, 1391 (2012).
\bibitem{nr} W. H. Press, S. A. Teukolsky, W. T. Vetterling, and B. P.
Flannery, \textsl{Numerical Recipes: The Art of Scientific Computing}
(Cambridge University Press, New York, 1986).

\bibitem{lapun} I. Shimada and T. Nagashima, Prog. Theor. Phys. 61 1605
(1979).
\end{thebibliography}
\end{document}